\newcommand{\nc}{\newcommand}
\nc{\beq}{\begin{equation}}   \nc{\eeq}{\end{equation}}
\nc{\bea}{\begin{eqnarray}}   \nc{\eea}{\end{eqnarray}}
\nc{\baa}{\begin{array}}      \nc{\eaa}{\end{array}}
\nc{\bit}{\begin{itemize}}    \nc{\eit}{\end{itemize}}
\nc{\ben}{\begin{enumerate}}  \nc{\een}{\end{enumerate}}
\nc{\bce}{\begin{center}}     \nc{\ece}{\end{center}}
\def\beqa{\begin{eqnarray}}
\def\eeqa{\end{eqnarray}}

\def\lsim{\mathrel{\raise.3ex\hbox{$<$\kern-.75em\lower1ex\hbox{$\sim$}}}}
\def\gsim{\mathrel{\raise.3ex\hbox{$>$\kern-.75em\lower1ex\hbox{$\sim$}}}}

\def\be{\beq}
\def\ee{\eeq}
\def\to{\rightarrow}

\def\PHYSICA #1 #2 #3 {{\sl Physica}~{\bf#1} (#3) #2}
\def\MPL #1 #2 #3 {{\sl Mod.~Phys.~Lett.}~{\bf#1} (#3) #2}
\def\NPB #1 #2 #3 {{\sl Nucl.~Phys.}~{\bf #1} (#3) #2}
\def\NPBPS #1 #2 #3 {{\sl Nucl.~Phys.~B~(Proc. Suppl.)}~{\bf #1} (#3) #2}
\def\PLB #1 #2 #3 {{\sl Phys.~Lett.}~{\bf #1} (#3) #2}
\def\PR #1 #2 #3 {{\sl Phys.~Rep.}~{\bf#1} (#3) #2}
\def\PRD #1 #2 #3 {{\sl Phys.~Rev.}~{\bf #1} (#3) #2}
\def\PRL #1 #2 #3 {{\sl Phys.~Rev.~Lett.}~{\bf#1} (#3) #2}
\def\RMP #1 #2 #3 {{\sl Rev.~Mod.~Phys.}~{\bf#1} (#3) #2}
\def\ZPC #1 #2 #3 {{\sl Z.~Phys.}~{\bf #1} (#3) #2}
\def\IJMP #1 #2 #3 {{\sl Int.~J.~Mod.~Phys.}~{\bf#1} (#3) #2}

\documentstyle[12pt,epsfig]{elsart}

\begin{document}  
\newlength{\captsize} \let\captsize=\small 
\newlength{\captwidth}                     

\begin{frontmatter}
\font\fortssbx=cmssbx10 scaled \magstep2
\hbox to \hsize{
$\vcenter{
\hbox{\fortssbx Universidade de S\~ao Paulo}
\hbox{\fortssbx Universidade Estadual de Campinas}
}$
\hfill
$\vcenter{
\hbox{\bf IFUSP-DFN/99-035}
\hbox{\bf hep-ph/9911470}
\hbox{November, 1999}
}$
}


\title{
Testing Flavor Changing Neutrino Interactions in 
 Long Baseline Experiments}

\author[USP,PUC]{A. M. Gago\thanksref{gago}},
\author[USP]{L. P. Freitas\thanksref{lfreitas}},
\author[UNICAMP]{O. L. G. Peres\thanksref{orlando}} and 
\author[USP]{R. Zukanovich Funchal\thanksref{zukanov}}

\address[USP]{Instituto de F\'\i sica, Universidade de S\~ao Paulo,
    C.\ P.\ 66.318, 05315-970\\  S\~ao Paulo, Brazil}

\address[PUC]{Secci\'on F\'{\i}sica, Departamento de Ciencias,
    Pontificia Universidad Cat\'{o}lica del Per\'{u}, 
    Apartado 1761, Lima, Per\'{u}}

\address[UNICAMP]{Instituto de F\'{\i}sica Gleb Wataghin, Universidade 
Estadual de Campinas -- UNICAMP, 13083-970 Campinas, Brazil}

\thanks[gago]{Email address: agago@charme.if.usp.br}
\thanks[lfreitas]{ Email address: lfreitas@charme.if.usp.br} 
\thanks[orlando]{Email address: orlando@ifi.unicamp.br} 
\thanks[zukanov]{Email address: zukanov@charme.if.usp.br}

\begin{abstract}
\noindent We have investigated the possibility of discerning 
mass from flavor changing neutrino interactions 
induced $\nu_\mu \to \nu_\tau$ oscillations  in the long baseline 
neutrino experiments K2K and MINOS. We have found that for virtually 
any value of the flavor conserving parameter $\epsilon^\prime$ it will be 
possible to, independently, distinguish these two mechanisms at K2K, 
if the flavor changing parameter $\epsilon$  is in the range
$\epsilon \gsim 0.77$, and at MINOS, if $\epsilon \gsim 0.2$. 
Moreover, if K2K measures a depletion of the expected $\nu_\mu$ flux 
then MINOS will either observe or discard completely flavor 
changing neutrino oscillations.
\end{abstract}
\end{frontmatter}


\section{Introduction} 
\label{sec:int1}

Although over three decades of solar neutrino experiments~\cite{solarexp} 
and one  decade of atmospheric neutrino data~\cite{sk,sk1} have confirmed, 
beyond any reasonable doubt, that neutrinos have indeed an oscillating
nature, the question about 
which is the dynamical mechanism responsible for such oscillations still 
remains an open one.   

We can find in the literature today a variety of different schemes~\cite{ourwork,ourwork1,lisi,kb97,FC-mass,decay,vacuo,msw,vepvacuum}, going from mass induced to 
gravitationally induced oscillations passing by decaying neutrinos, that can successfully account for the 
solar and/or atmospheric neutrino results. 
It is therefore important to investigate the possibility of 
distinguishing different solutions in experiments where neutrino beams are 
manufactured by  accelerators in the Earth. 
These experiments have the advantage that one can control 
what is being produced (neutrino flux, flavor, energy distribution) as well 
as what is being detected. We therefore believe that they must provide the 
definite proof, not only of neutrino conversion, through the observation of 
neutrino flavor appearance or disappearance, but also of the dynamical 
nature of the oscillation process.  Besides, it is always important to 
have an independent check of the operative neutrino oscillation mechanism.

It has been recently suggested~\cite{all} that flavor changing neutrino 
interactions (FCNI) could induce $\nu_\mu \to \nu_\tau$ oscillations that 
would explain rather well the sub-GeV and multi-GeV data reported by 
the Super-Kamiokande atmospheric neutrino experiment. Since then some 
controversy about the quality of this solution when one also includes the 
upward-going muon data in the FCNI analyses has come 
up~\cite{lilu,fcv,taup99}. 
In any case, regardless of the discussion on the goodness of the fit to the 
atmospheric neutrino data, the proposed FCNI solution selects a parameter 
space region that will be well in the reach of the forthcoming long baseline 
experiments. We will in what follows refer to this type of oscillation simply 
as flavor changing induced oscillation (FCIO).

It has been pointed out in Ref.~\cite{bg} that a fine tuning of model 
parameters would be necessary in order to reconcile the FCIO solution 
to the atmospheric neutrino problem with current experimental limits on 
lepton number violation coming from lepton decays. However 
it will be up to nature to confirm or disclaim our theoretical 
preferences and prejudices.
Having this in mind we have investigated the possibility of using the  
long baseline K2K~\cite{kek} and MINOS~\cite{minos} neutrino accelerator 
experiments in order to discriminate mass induced  $\nu_\mu \to \nu_\tau$ 
oscillations (MIO) from FCIO. Many authors have suggested and discussed the 
physical capabilities of these experiments~\cite{lbl-old,pakvasa}. 
Both of them will have neutrino beams that will travel through a
certain amount of Earth matter before reaching  the corresponding
neutrino detector, making them specially suited to probe the FCNI mechanism. 

The outline of the paper is as follows. In Section 2, we briefly revise 
the two oscillation formalisms and recall, in each case, the distinct 
features of the $\nu_\mu \to \nu_\tau$ oscillation  probability. 
In Section 3 we define the  K2K and  MINOS observables that we have used 
in our analysis and  describe in detail how we have estimated their 
values as a function of the free oscillation parameters.  
In Section 4 we present and discuss our results for K2K and MINOS.
Our conclusions are presented in Section 5.

\section{Review of the MIO and FCIO Formalisms} 
\label{sec:sec2}

\subsection{Mass Induced Oscillations}

Maki, Nakagawa and Sakata~\cite{mns} were the first to argue that if 
neutrinos have mass their states that are created and detected by weak 
interactions may not be the eigenstates of propagation and therefore 
neutrino flavor oscillations can occur.  
This is the simplest neutrino oscillation mechanism 
and perhaps the most likely one to take place in nature.

We will consider that oscillations between $\nu_\mu \to \nu_\tau$ 
can occur  in a two family scenario with two massive neutrinos. In this case 
neutrino mass eigenstates and flavor eigenstates  can be related 
by a Cabibbo-like mixing matrix, that can be parameterized by a single  
angle. The neutrino evolution Hamiltonian in vacuum is 
rather trivial and the corresponding equations can be solved analytically 
to compute  the oscillation probability between the two flavor 
eigenstates $\nu_{\mu} \rightarrow \nu_\tau$,
$P_{\nu_{\mu} \rightarrow \nu_\tau}^{\mbox{\scriptsize mass}}$, to obtain the well known formula:

\be
P_{\nu_{\mu} \rightarrow \nu_\tau}^{\mbox{\scriptsize mass}} \equiv P_{\nu_{\mu} \rightarrow \nu_\tau} \left( \sin^2 2 \theta ,\Delta m^2, E_\nu, L \right) 
= \sin^2  (2 \theta) \, \sin^2 \left( \pi \frac{L}{L^{\mbox{\scriptsize m}}_{\mbox{\scriptsize osc}}} \right),
\label{osc}
\ee

where $\theta$ is the mixing angle, the two non-degenerate neutrinos have 
mass $m_1$ and $m_2$, $\Delta m^2 = |m_1^2-m_2^2|$ is   
measured in eV$^2$, $L$ is the distance between the neutrino source and the 
detector in km, and $E_\nu$ the neutrino energy, in GeV. The neutrino 
oscillation length, which is also measured in km and grows linearly with 
the neutrino energy, is defined as

\be
L^{\mbox{\scriptsize m}}_{\mbox{{\scriptsize osc}}}= \frac{ \pi E_\nu}{1.27 \Delta m^2}.
\label{osclen}
\ee

We have here two free parameters, $\sin^2 2\theta$ and $\Delta m^2$.
Inspecting Eq.~(\ref{osc}) we see that the maximum of the conversion 
probability, for a constant amplitude, happens when 
$L/L^{m}_{\mbox{\scriptsize osc}}=1/2$. This condition is satisfied
for fixed $L$  and an averaged neutrino energy $\langle E_{\nu} \rangle$ for   

\begin{equation}
\Delta m^2_*= \frac{\pi}{2\times 1.27}\left(
\frac{\langle E_{\nu} \rangle}{L}\right).
\label{2mm}
\end{equation}

\subsection{Flavor Changing Induced Oscillations}

The fact that FCNI can induce neutrino oscillations in matter
was first investigated by Wolfenstein~\cite{lw} who pointed out that 
interactions in a medium modify the dispersion relations of particles 
traveling through. Wolfenstein effect generates quantum phases in the time 
evolution of phenomenological neutrinos eigenstates which consequently can 
oscillate. 

In a two-flavor mixing scheme the presence of flavor changing 
neutrino-matter interactions implies a
non-trivial structure for the neutrino evolution Hamiltonian in matter,
even if massless neutrinos and no mixing in the vacuum is assumed.  The
evolution equations describing  the $\nu_\mu \rightarrow \nu_\tau$
transitions are  given by~\cite{gmp}:

\be   
i{\displaystyle\frac{d}{dr}} \left( \begin{array}{c} \nu_\mu \\ \nu_\tau \end{array} \right) 
=  \sqrt{2}\,G_F \left( \begin{array}{cc} 0 &  \epsilon^f n_f(r)
\\ \epsilon^f n_f(r)& \epsilon '^f n_f(r) 
\end{array} \right)
\left( \begin{array}{c} \nu_\mu \\ \nu_\tau  \end{array} \right),
\label{motion} 
\ee
where $\nu_\mu \equiv \nu_\mu (r)$ and $\nu_\tau \equiv \nu_\tau (r)$, are the 
probability amplitudes to find these neutrinos at a distance $r$ from their 
creation position, $\sqrt{2}\,G_F n_f(r) \epsilon^f$ is the 
flavor-changing $\nu_\mu + f \to \nu_\tau +f$ forward scattering amplitude 
with the interacting fermion $f$ (electron, $d$ or $u$ quark) 
and $\sqrt{2}\,G_F n_f(r) \epsilon '^f$ is the difference 
between the flavor diagonal $\nu_\mu - f$ and $\nu_\tau - f$ elastic 
forward scattering amplitudes, with $n_f(r)$ being the number density of the 
fermions which induce these processes. From now on we will consider FCNI 
only with a single fermion type and drop the label $f$ attached to 
$\epsilon$ and  $\epsilon^\prime$.

For  constant matter density Eq.~(\ref{motion}) can be analytically solved
to give a conversion probability,  
$P_{\nu_{\mu} \rightarrow \nu_\tau}^{\mbox{\scriptsize FCNI}}$, that can 
be written as 

\be
P_{\nu_{\mu} \rightarrow \nu_\tau}^{\mbox{\scriptsize FCNI}} \equiv P_{\nu_\mu \rightarrow \nu_\tau} 
\left( \epsilon ,\epsilon^{\prime}, L \right) = 
\frac{4 \epsilon^2}{4 \epsilon^2 + {\epsilon^{\prime}}^2} \sin^2 
\left( \pi \frac{L}{L^{\mbox{\scriptsize f}}_{\mbox{\scriptsize osc}}} \right),
\label{FCC}
\ee
where $L$ is the neutrino flight length from the production source to the 
detector and $L^{\mbox{\scriptsize f}}_{\mbox{\scriptsize osc}}$ is the 
oscillation length, both measured in km.
 $L^{\mbox{\scriptsize f}}_{\mbox{\scriptsize osc}}$, which in contrast to 
Eq.~(\ref{osc}), does not depend on the neutrino energy, can be written 
explicitly as follows:

\be
L^{\mbox{\scriptsize f}}_{\mbox{\scriptsize osc}} = 
2707.4 \times \frac{3}{C} \times 
\frac{\mbox{(2 mol/cm}^3) }{n_e} 
\frac{1}{\sqrt{4 \epsilon^2 + \epsilon^{\prime 2}}}, 
\ee
where $C = 3$,  for FCNI with $u$- or $d$-quarks, $C = 1$, for FCNI with electrons, 
and $n_e$ is the Earth's electron density in mol/cm$^3$. For $n_e=2$ mol/cm$^3$
and FCNI only with $u$ or $d$ quarks 
$L^{\mbox{\scriptsize f}}_{\mbox{\scriptsize osc}} = $ 2707.4 km.
We have in this scheme also two free parameters, $\epsilon$ and $\epsilon^\prime$.
In this paper we will only consider FCNI involving $d$-quarks.
If we were to consider electrons as the interacting fermions, instead of 
$d$-quarks, this would mean a simple rescaled of our results by a factor 
three and so there is no need to show this explicitly here. 

In analogy to the discussion made for MIO  it is instructive 
to point out the condition that has to be satisfied in order to 
the conversion probability to be at a maximum for the FCIO
mechanism. If we impose $L/L^f_{\mbox{osc}}= 1/2$ then

\begin{equation}
\sqrt{4 \epsilon^2_*+ \epsilon^{\prime 2}_*}=
 \left(\frac{2707.4}{2\times L}\right)
\left(\frac{3}{C}\right)\left(\frac{\mbox{(2 mol/cm}^3) }{n_e}\right),
\end{equation}

and if $\epsilon^{\prime}$ and $\epsilon$ have comparable sizes 

\begin{equation}
\epsilon_* \sim
 \left(\frac{2707.4}{4\times L}\right)
\left(\frac{3}{C}\right)\left(\frac{\mbox{(2 mol/cm}^3) }{n_e}\right).
\label{nn}
\end{equation}

\section{Description of the MIO and FCIO analysis for K2K and MINOS} 
\label{sec:sec3}

\subsection{K2K Experiment}

K2K~\cite{kek} is a long baseline experiment 
were a $\nu_\mu$ beam is produced by the Japanese 
KEK accelerator, driven through a certain amount of the Earth's crust 
before reaching the K2K detector. 

In our approach we only consider $\nu_\mu \to \nu_\tau$ transitions, 
disregarding as completely negligible the channel $\nu_\mu \to \nu_e$, and 
since the K2K experiment is unable, due to its range in energy, to observe 
$\nu_\tau$ production, the only information which is valuable to us is a  
possible measurement of the reduction of the $\nu_\mu$ flux. 
For this reason, we have computed the mean oscillation probability 
$\langle P_{\nu_{\mu} \rightarrow \nu_{\tau}} ( \sin^2 (2 \theta) ,\Delta m^2 ) \rangle$ that can be measured by K2K as a function of the MIO free parameters $\sin^2 2\theta$ and $\Delta m^2$ defined as 

\be
\langle P_{\nu_{\mu} \rightarrow \nu_{\tau}} ( \sin^2 (2 \theta) ,\Delta m^2 ) \rangle= \displaystyle \frac{\int \int dx \, dE_\nu \, h(E_\nu) \, 
f(x,L_{\mbox{\scriptsize K2K}}) \, 
P_{\nu_{\mu} \rightarrow \nu_{\tau}}^{\mbox{\scriptsize mass}}} {\int h(E_\nu) dE_\nu},
\label{kekprob}
\ee
where $L_{\mbox{\scriptsize K2K}}=250$ km, $h(E_\nu)$ is the predicted 
neutrino energy spectrum for $\nu_{\mu}$ in the far detector 
that can be found in the Ref.~\cite{kek} and  
$P_{\nu_{\mu} \rightarrow \nu_\tau}^{\mbox{\scriptsize mass}}$ is the 
probability showed in  Eq.~(\ref{osc}). 
In order to take into account the uncertainties in the distance $L$ we 
use in Eq.~(\ref{kekprob}) the following Gaussian smearing function $f(x,L)$: 

\begin{equation}
f(x,L)=\frac{1}{\sqrt{2 \pi} \sigma}\exp \left[-\frac{(x - L)^2}{2 \sigma^2} 
 \right], 
\label{smear}
\end{equation}
where we have assumed $\sigma= 0.05 \, L_{\mbox{\scriptsize K2K}}$.

In analogy, we have obtained the mean oscillation probability 
$\langle P_{\nu_{\mu} \rightarrow \nu_{\tau}} \left( \epsilon ,\epsilon^{\prime} \right) \rangle$ for FCIO as a function of the free parameters $\epsilon$ 
and $\epsilon^\prime$ defined by

\be
\langle P_{\nu_{\mu} \rightarrow \nu_{\tau}}
\left( \epsilon ,\epsilon^{\prime} \right) \rangle = 
\int dx \, f(x,L_{\mbox{\scriptsize K2K}}) \ 
P_{\nu_{\mu} \rightarrow \nu_{\tau}}^{\mbox{\scriptsize FCNI}}, 
\label{probfc}
\ee
where $f(x,L_{\mbox{\scriptsize K2K}})$ is the smearing function given 
in Eq.\ (\ref{smear}) and $P_{\nu_{\mu} \rightarrow \nu_{\tau}}^{\mbox{\scriptsize FCNI}}$ is the 
probability given in Eq. (\ref{FCC}). 
We have used in our calculations the Earth's electron density profile for 
K2K  given in Ref. \cite{sato}, from this profile one can calculate that 
the mean electron density will be $n_e = 2.35$~mol/cm$^3$. 
We use this mean electron density to compute the 
oscillation probability. Note that since 
the FCIO is energy independent, we do not need to take the average of  
Eq.~(\ref{probfc}) over the neutrino energy spectrum.

\subsection{MINOS Experiment}

The MINOS experiment~\cite{minos} is a part of the Fermilab NuMI Project.
The neutrinos which constitute the MINOS beam will be  the result of the  
decay of pions and kaons that will be produced by the 120 GeV proton high 
intensity beam extracted from the Fermilab Main Injector. There will be two MINOS 
detectors, one located at Fermilab (the near detector) and  another located 
in the Soudan mine in Minnesota, about 730 km away (the far detector). 
MINOS will be thus a $L_{\mbox{\scriptsize MIN}}=730$ km long baseline experiment.

According to Ref.~\cite{minostec}, MINOS will be able to measure 
independently the rates and the energy spectra for neutral current
(nc) and charged current (cc) reactions. About 3000 $\nu_\mu$
cc-events/kt/year are expected in the MINOS far detector for the highest 
energy configuration. We will compute here the ratio 
$R_{{\mbox{\scriptsize nc}/{\mbox{\scriptsize cc}}}}$ that should be 
expected at MINOS for MIO and FCIO as a function of 
the respective free parameters. This ratio has the advantage that it does not 
require the understanding of the relative fluxes 
at the near and far detectors, it is also quite sensitive to oscillations 
since when they occur not only cc-events are depleted but nc-events are 
enhanced.

In the MIO or FCIO hypothesis this ratio can be written as~\cite{numi}:

\be
R_{{\mbox{\scriptsize nc}/{\mbox{\scriptsize cc}}}}^{\mbox{\scriptsize osc}} = \frac{\int dE_\nu \, N_{\mbox{\scriptsize nc}}^{\mbox{\scriptsize osc}}(E_\nu)} {\int dE_\nu \, N_{\mbox{\scriptsize cc}}^{\mbox{\scriptsize osc}}(E_\nu)},
\label{ratio1}
\ee 

where 
\be 
 N_{\mbox{\scriptsize cc}}^{\mbox{\scriptsize osc}}(E_\nu) = N^{\mbox{\scriptsize no-osc}}_{\mbox{\scriptsize cc}}(E_\nu)(1-P_{\nu_\mu \rightarrow \nu_\tau})+ N^{\mbox{\scriptsize no-osc}}_{\mbox{\scriptsize cc}}(E_\nu) \eta(E_\nu) B P_{\nu_\mu \rightarrow \nu_\tau}, 
\label{ncc}
\ee

describes the two possible ways that a muon can be produced; the first term 
representing the contribution of surviving $\nu_\mu$ and the second the 
contribution of $\tau \rightarrow \nu_\mu \nu_\tau \mu$ decays, from taus  
generated by $\nu_\tau$ interactions in the detector after  
$\nu_\mu \to \nu_\tau$ conversion and 
 
\be 
 N_{\mbox{\scriptsize nc}}^{\mbox{\scriptsize osc}}(E_\nu) = N^{\mbox{\scriptsize no-osc}}_{\mbox{\scriptsize nc}}(E_\nu) + N^{\mbox{\scriptsize no-osc}}_{\mbox{\scriptsize cc}}(E_\nu) \eta(E_\nu) \, (1-B) P_{\nu_\mu \rightarrow \nu_\tau}, 
\label{nnc}
\ee

is the nc contribution with

\be
\eta(E_\nu) = \frac{\sigma^{\mbox{\scriptsize cc}}_{\nu_\tau}(E_\nu)}{\sigma^{\mbox{\scriptsize cc}}_{\nu_\mu}(E_\nu)},
\label{eta}
\ee

where $B$ is 0.18, the branching ratio for $\tau$ leptonic decay,
$N^{\mbox{\scriptsize no-osc}}_{\mbox{\scriptsize cc}}$ is the energy 
spectrum for $\nu_\mu$ cc-events in the MINOS far detector in the case 
of no oscillation~\cite{minostec}. The high-energy wide-band beam was assumed and the 
spectrum  has already been smeared by the expected detector
resolution~\cite{minostec}.  From this spectrum we have inferred
$N^{\mbox{\scriptsize no-osc}}_{\mbox{\scriptsize nc}}$, the expected
energy spectrum for $\nu_\mu$ nc-events in the MINOS far detector in the case of no oscillation, 
using the approximation:

\be
 N_{\mbox{\scriptsize nc}}^{\mbox{\scriptsize no-osc}}(E_\nu) =  
N_{\mbox{\scriptsize cc}}^{\mbox{\scriptsize no-osc}}(E_\nu)
\frac{\sigma^{\mbox{\scriptsize nc}}_{\nu_\mu}(E_\nu)}{\sigma^{\mbox{\scriptsize cc}}_{\nu_\mu}(E_\nu)}.
\ee

The cross sections $\sigma^{\mbox{\scriptsize cc}}_{\nu_\mu}$, 
$\sigma^{\mbox{\scriptsize cc}}_{\nu_\tau}$ and 
$\sigma^{\mbox{\scriptsize nc}}_{\nu_\mu}$ 
were taken from Refs. \cite{phd,dutta,okumura,nc} and 
the conversion probability, $P_{\nu_\mu \rightarrow \nu_\tau}$,  
used for the MIO case is defined by 

\be
P_{\nu_\mu \to \nu_\tau} = \int dx  f(x,L_{\mbox{\scriptsize MIN}}) \,
P_{\nu_{\mu} \rightarrow \nu_{\tau}}^{\mbox{\scriptsize mass}},
\label{mminos}
\ee

and for FCIO case by

\be
P_{\nu_\mu \to \nu_\tau} = 
\langle P_{\nu_{\mu} \rightarrow \nu_{\tau}}
\left( \epsilon ,\epsilon^{\prime} \right) \rangle = 
\int dx \, f(x,L_{\mbox{\scriptsize MIN}}) \ 
P_{\nu_{\mu} \rightarrow \nu_{\tau}}^{\mbox{\scriptsize FCNI}},
\label{fminos} 
\ee

where $f(x,L_{\mbox{\scriptsize MIN}})$ is the smearing function given 
in Eq.\ (\ref{smear}) with $\sigma= 0.05 \, L_{\mbox{\scriptsize MIN}}$
and $P_{\nu_{\mu} \rightarrow \nu_{\tau}}^{\mbox{\scriptsize mass}}$, 
$P_{\nu_{\mu} \rightarrow \nu_{\tau}}^{\mbox{\scriptsize FCNI}}$ are 
the probabilities given in Eqs.~(\ref{osc}) and (\ref{FCC}) respectively. 

In all the calculations involving FCIO we 
have assumed for MINOS a constant density of $n_e = 2.80$~mol/cm$^3$, 
which is the typical value for rock~\cite{lipari1,priv}.

\section{Presentation and Discussion of Results} 
\label{sec:fin}

We  have computed the region in the $\epsilon \times \epsilon^\prime$ 
plane that will be attainable by K2K and by MINOS.
In order to do this, we have compared the probabilities given in 
Eqs.~(\ref{probfc}) and (\ref{fminos}) with the sensitivity limit, 
$P^{\mbox{\scriptsize lim}}_{\nu_{\mu} \rightarrow \nu_\tau}$, for each one 
of these experiments and selected the pairs of 
values ($\epsilon$,$\epsilon^\prime$) that satisfied the condition:

\begin{equation}
\langle P_{\nu_{\mu} \rightarrow \nu_\tau} 
( \epsilon ,\epsilon^{\prime}) \rangle
\ge  P^{\mbox{\scriptsize lim}}_{\nu_{\mu} \rightarrow \nu_\tau}, 
\label{plim}
\end{equation}
where $\langle P_{\nu_{\mu} \rightarrow \nu_\tau} ( \epsilon ,\epsilon^{\prime}) \rangle$  and $P^{\mbox{\scriptsize lim}}_{\nu_{\mu} \rightarrow \nu_\tau}$
are  respectively given by Eq.~(\ref{probfc}) and equal to 0.25 in the case 
of  K2K and  given by Eq.~(\ref{fminos}) and equal to 0.010 in the 
case of MINOS.

\begin{figure}
\begin{center}
\parbox[c]{6.5in}
{\mbox{\qquad\epsfig{file=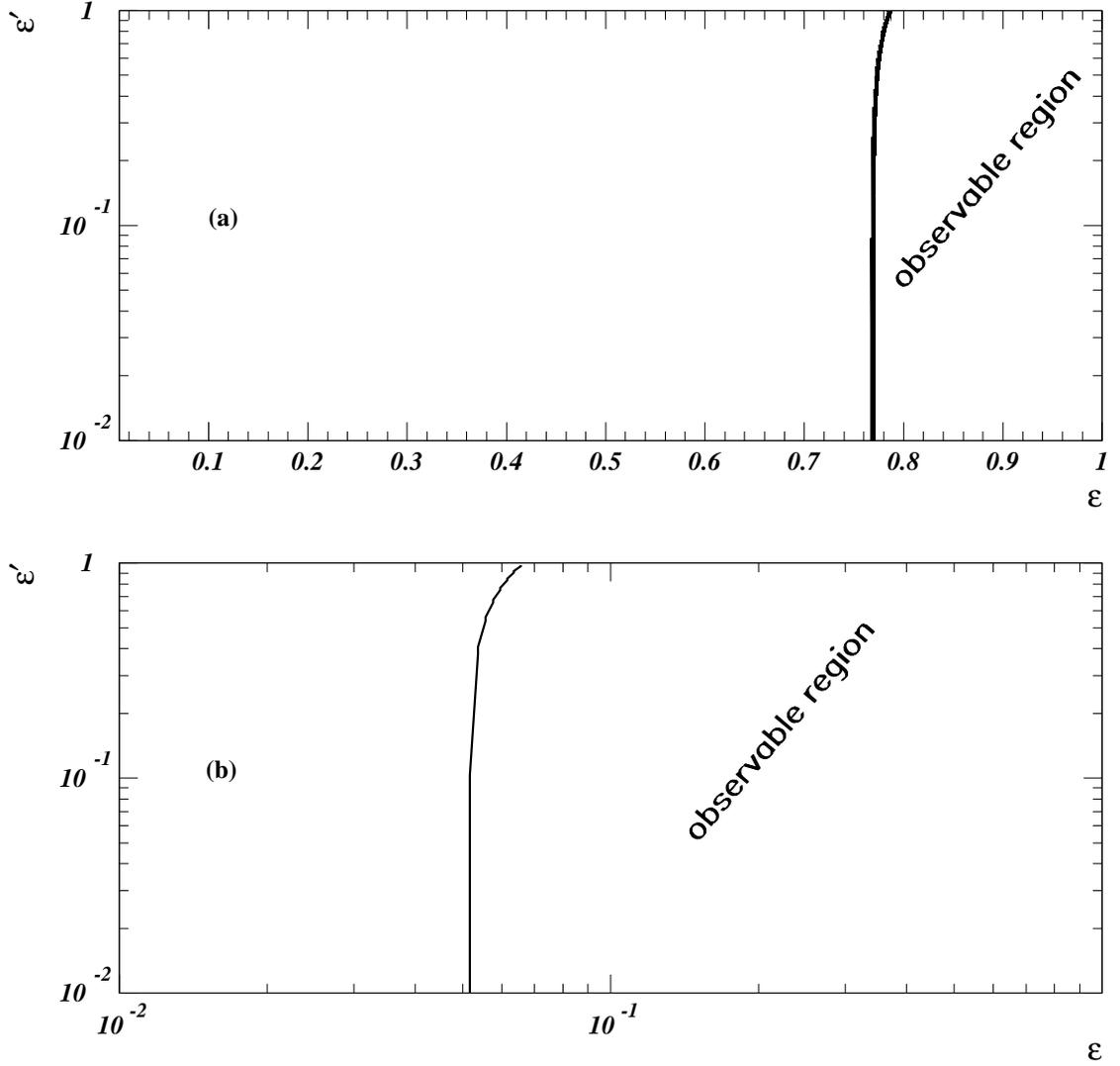,width=\linewidth}}}
\end{center}
\caption{Region in the $\epsilon \times \epsilon^{\prime}$ plane 
that will be observable in (a) K2K  and  (b) MINOS 
in the case of FCNI induced $\nu_{\mu} \rightarrow \nu_\tau$ oscillations.}   
\label{fig1}
\end{figure}

We display in Figs.~(\ref{fig1})(a) and (b) the observable values for the 
parameters $\epsilon$ and $\epsilon^{\prime}$ obtained by applying the 
condition given by Eq.~(\ref{plim}) to K2K and MINOS respectively.
We see that for K2K $\epsilon \gsim 0.77$ for any value of $\epsilon^\prime$, 
while for MINOS one can reach much smaller values since 
$\epsilon \gsim 5 \times 10^{-2}$.  
In fact we can see from Figs.~(\ref{fig1})(b) that virtually all values of 
$\epsilon$ that are allowed by the FCIO analysis of the Super-Kamiokande 
data~\cite{all} can in principle be tested by the MINOS long baseline 
experiment.

\begin{figure}[htb]
\begin{center}
\parbox[c]{6.5in}
{\mbox{\qquad\epsfig{file=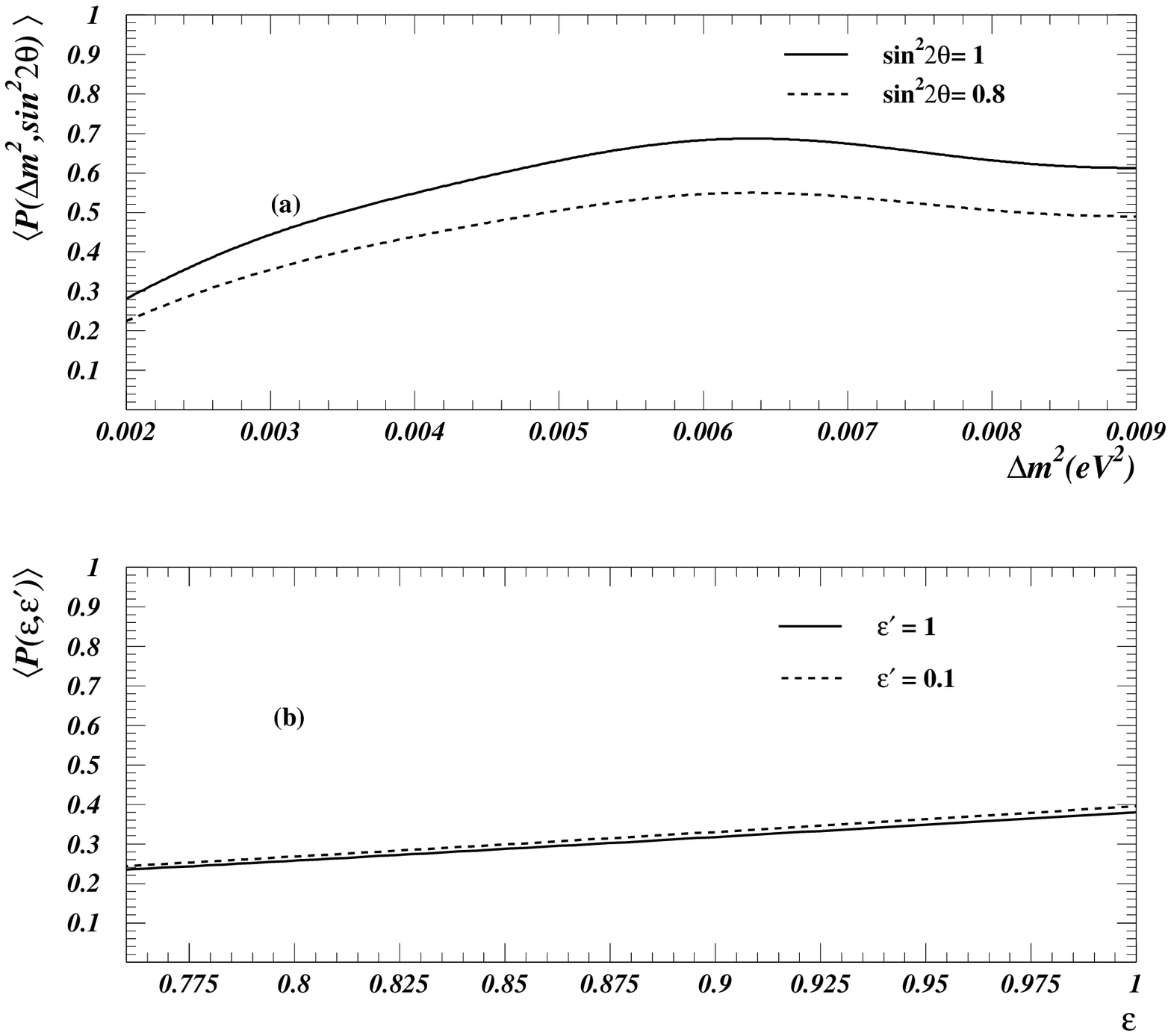,width=\linewidth}}}
\end{center}
\caption{The averaged probability of $\nu_\mu \to \nu_\tau$ conversion 
for the K2K experiment (a) in the case of  MIO as a function 
of $\Delta m^2$ and (b) in the case of FCIO as a function of $\epsilon$. 
}
\label{fig2}
\end{figure}

\subsection{K2K}

In Figs.\ref{fig2} we show the averaged probabilities  
$\langle P_{\nu_{\mu} \rightarrow \nu_{\tau}} ( \sin^2 (2 \theta)
,\Delta m^2 ) \rangle$ for MIO  and 
$\langle P_{\nu_{\mu} \rightarrow \nu_{\tau}}
\left( \epsilon ,\epsilon^{\prime} \right) \rangle$ for 
FCIO that can be measured by the K2K experiment. 
In Fig.\ref{fig2}(a) the averaged probabilities for mass induced oscillation 
with $\sin^2 (2\theta)$=0.8 and 1.0 are shown as a function of $\Delta m^2$. 
These two curves give the maximum and minimum probabilities 
as a function of $\Delta m^2$ that can be reached if the oscillation 
parameters are chosen inside the 90\% C. L. allowed region for 
Super-Kamiokande solution to the atmospheric neutrino problem~\cite{SKconf}, 
i.e. $0.8 \le \sin^2 (2 \theta)\le 1.0$ and 
$2.0 \times 10^{-3} \mbox{ eV}^2 \le \Delta m^2 \le 9.0 \times 10^{-3} 
\mbox{ eV}^2$.
We note that in the range of parameters considered here the averaged 
oscillation probability of K2K can vary from a minimum value around 0.25, 
which is the sensitivity  limit, to a maximal value around 0.69, which  
corresponds to $\Delta m^2_{*} \sim \displaystyle \langle E_\nu
\rangle/L=6 \times 10 ^{-3}$ eV$^2$, just at the sensitivity of K2K,
as can be expected from Eq.~(\ref{2mm}) for an averaged 
neutrino energy of  about 2 GeV.

In the Fig.~\ref{fig2}(b) we present two curves for the averaged 
probability in the case of FCIO, as a function of 
 $\epsilon$ in the observable region of the K2K experiment. These curves 
give the maximum and minimum averaged probabilities as a 
function of $\epsilon$ that can be reached if $0.1 \le \epsilon^\prime \le 1$.
We observe that for this range of $\epsilon^{\prime}$ these curves are almost 
indistinguishable  one from the other.  This reflects the fact that the 
probability given in Eq.~(\ref{FCC}) can be written, when the 
argument of the sine is small, which is just the case for K2K 
($L_{\mbox{\scriptsize K2K}}=250$ km), simply as

\begin{equation}
P_{\nu_{\mu} \rightarrow \nu_\tau}^{\mbox{\scriptsize FCNI}}  \sim 
\frac{4 \epsilon^2}{4 \epsilon^2 + {\epsilon^{\prime}}^2} 
\left[
 \pi \left(\frac{L}{2707.4}\right)\left(\frac{C}{3}\right)
\left(\frac{n_e}{\mbox{2 mol/cm}^3 }\right)
\sqrt{4 \epsilon^2 + {\epsilon^{\prime}}^2}
\right]^2,
\label{FCC-new1}
\end{equation} 

which is independent of $\epsilon^\prime$ and grows proportionally to 
$\epsilon^2$. This is the reason why we have chosen to present in 
Fig.~\ref{fig2}(b) the averaged probability as a function of $\epsilon$ 
instead of $\epsilon^\prime$, which would be a more natural choice to 
compare with Fig.~\ref{fig2}(a). If FCIO are to be observed in K2K  
the averaged conversion probability will lay in the range 
0.25-0.4. 

If K2K observes a depletion of the $\nu_\mu$ flux their experimental  result 
can be translated into a certain  allowed region of averaged probability in 
Figs.~\ref{fig2}. It is possible that this region will be far enough 
from 0.4, the maximum allowed probability by FCIO, so that K2K may be able 
to rule out FCIO in the entire observable parameter 
region given in Fig.~\ref{fig1}(a). On the other hand if this allowed region 
turns out to be statistically compatible with  values in the range 0.25-0.4 
this observable  on its own will not be conclusive since both type of 
oscillations will be able to explain the data.
Nevertheless  K2K can measure  the energy spectrum of $\nu_\mu$
cc-events,  
thus we further investigated to what extent one could use this 
information to disentangle the two mechanisms of oscillation considered here.
   
\begin{figure}
\begin{center}
\parbox[c]{6.5in}
{\mbox{\qquad\epsfig{file=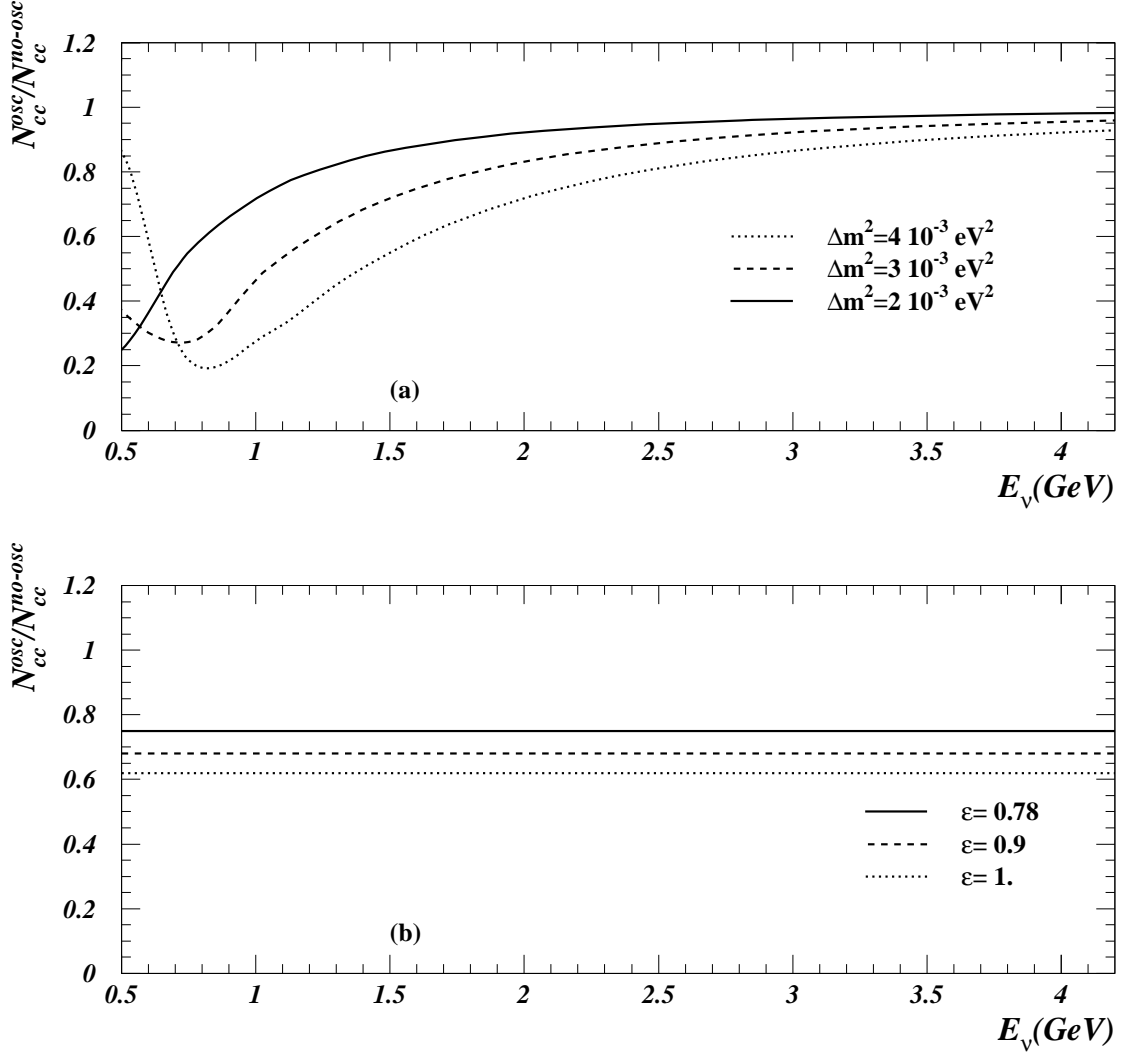,width=\linewidth}}}
\end{center}
\caption{Ratio of the energy spectrum of $\nu_\mu$ cc-events  for (a) 
MIO with $\sin^2 (2 \theta)=1.0$ and (b) FCIO with $\epsilon^\prime=1.0$
over the expected energy spectrum in the absence of neutrino oscillations for K2K.}
\label{fig3}
\end{figure}

In Figs.~\ref{fig3} we show the ratio, $N_{\mbox{\scriptsize cc}}^{\mbox{\scriptsize osc}}/N_{\mbox{\scriptsize cc}}^{\mbox{\scriptsize no-osc}}$, 
of the energy spectrum of $\nu_\mu$ cc-events in the hypothesis of oscillation 
over the expected energy spectrum without oscillation.
In Fig.~\ref{fig3}(a) MIO is considered with  
maximal oscillation amplitude ($\sin^2(2\theta)=1.0$) 
and three different values of $\Delta m^2$ in the range where the measured 
averaged oscillation probability cannot be conclusive. 
In Fig.~\ref{fig3}(b) we show a plot for FCIO, 
for $\epsilon^\prime=1.0$ and three different values of $\epsilon$. 
While MIO cause a pronounced spectral distortion 
at low energy, FCIO, which are energy independent, 
do not. According to the K2K detector simulation~\cite{oyama} they will 
be able to determine the neutrino energy in the region 
$E_\nu = 0.5 \sim 3.0$ GeV with better precision than 10\% 
so that we can expect that they will have enough accurate data points in the 
low energy region ($E_\nu \lsim  1.5$~GeV) to carefully examine possible  
spectral distortions and perhaps reach a definite conclusion about 
which is the mechanism responsible for $\nu_\mu \to \nu_\tau$ conversion 
at K2K.   

\subsection{MINOS}

\begin{figure}[htb]
\centering\leavevmode
\epsfxsize=520pt
\hglue -1.0cm
\epsfbox{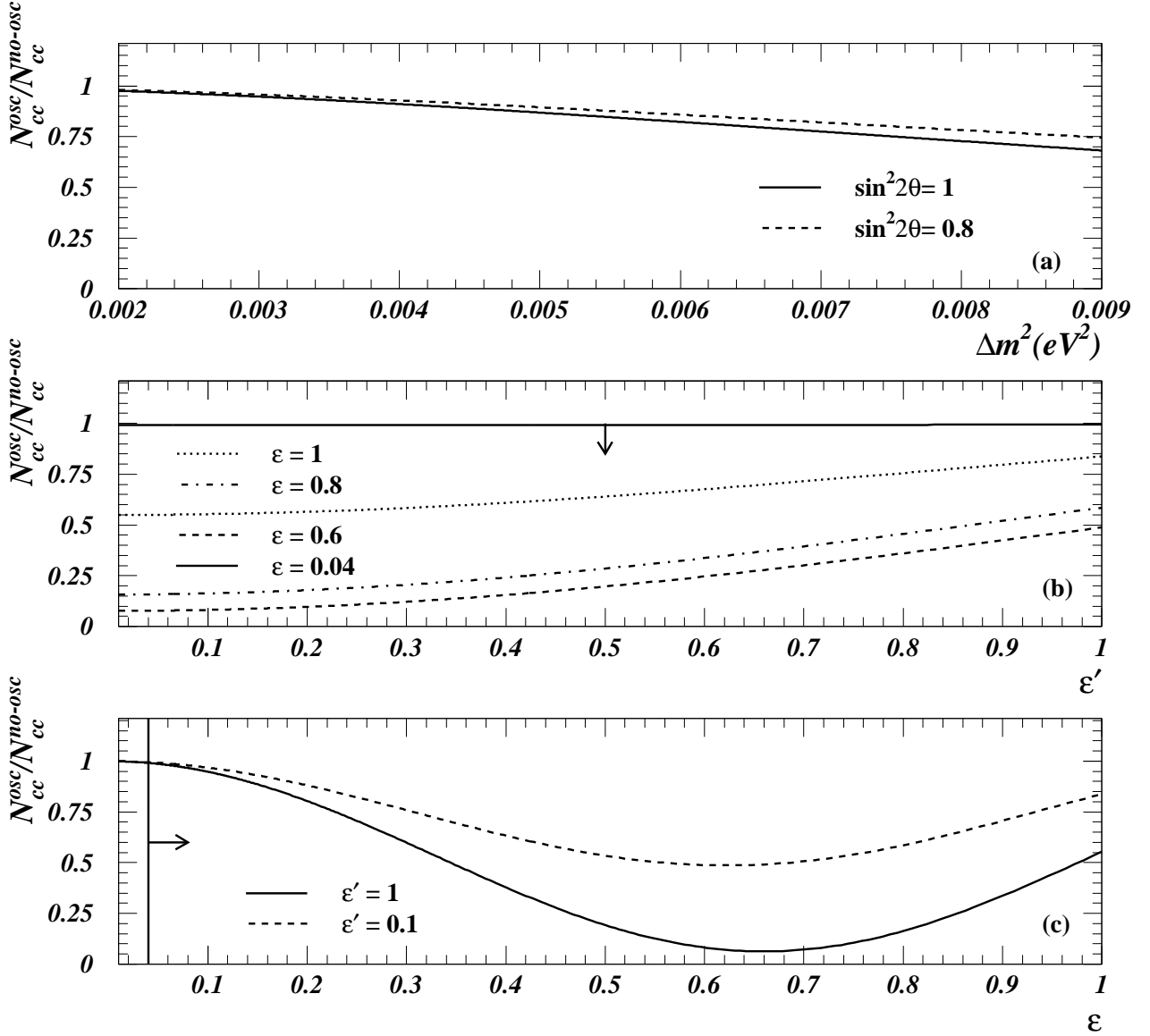}
\vglue -1.5cm
\caption{Ratio of the number of $\nu_\mu$ cc-events for (a) MIO as a function of 
$\Delta m^2$, (b) FCIO as a function of $\epsilon^\prime$ and (c) FCIO
as a function of $\epsilon$ over the number of $\nu_\mu$ cc-events  in 
the case of no oscillation. The solid line with the arrow in (b) and (c) marks 
the sensitivity of MINOS.}
\label{fignew}
\end{figure}

We show in Figs.~\ref{fignew}(a)-(c) the behavior of the ratio 
$N_{\mbox{\scriptsize cc}}^{\mbox{\scriptsize
osc}}/N_{\mbox{\scriptsize cc}}^{\mbox{\scriptsize no-osc}}$, which is 
essentially the $\nu_\mu$ survival probability for MINOS, for 
the MIO and FCIO mechanisms. In Fig.~\ref{fignew}(a) we see that 
this ratio for MIO is not very affected by $\sin^2 (2\theta)$, 
in the range allowed by the Super-Kamiokande experiment, 
and is always above $\sim$ 0.7 in the range of $\Delta m^2$ 
we have considered. On the other hand it depends rather strongly 
on the values of $\epsilon$ and of $\epsilon^\prime$ as 
can be seen in Figs.~\ref{fignew}(b) and (c). In particular for some 
values of $\epsilon$ there is a very strong suppression, 
when the conversion probability is close to its maximum, nevertheless
the ratio is never zero due to the contribution of $\tau$ decays as
shown in Eq.~(\ref{ncc}).

\begin{figure}[htb]
\begin{center}
\parbox[c]{6.5in}
{\mbox{\qquad\epsfig{file=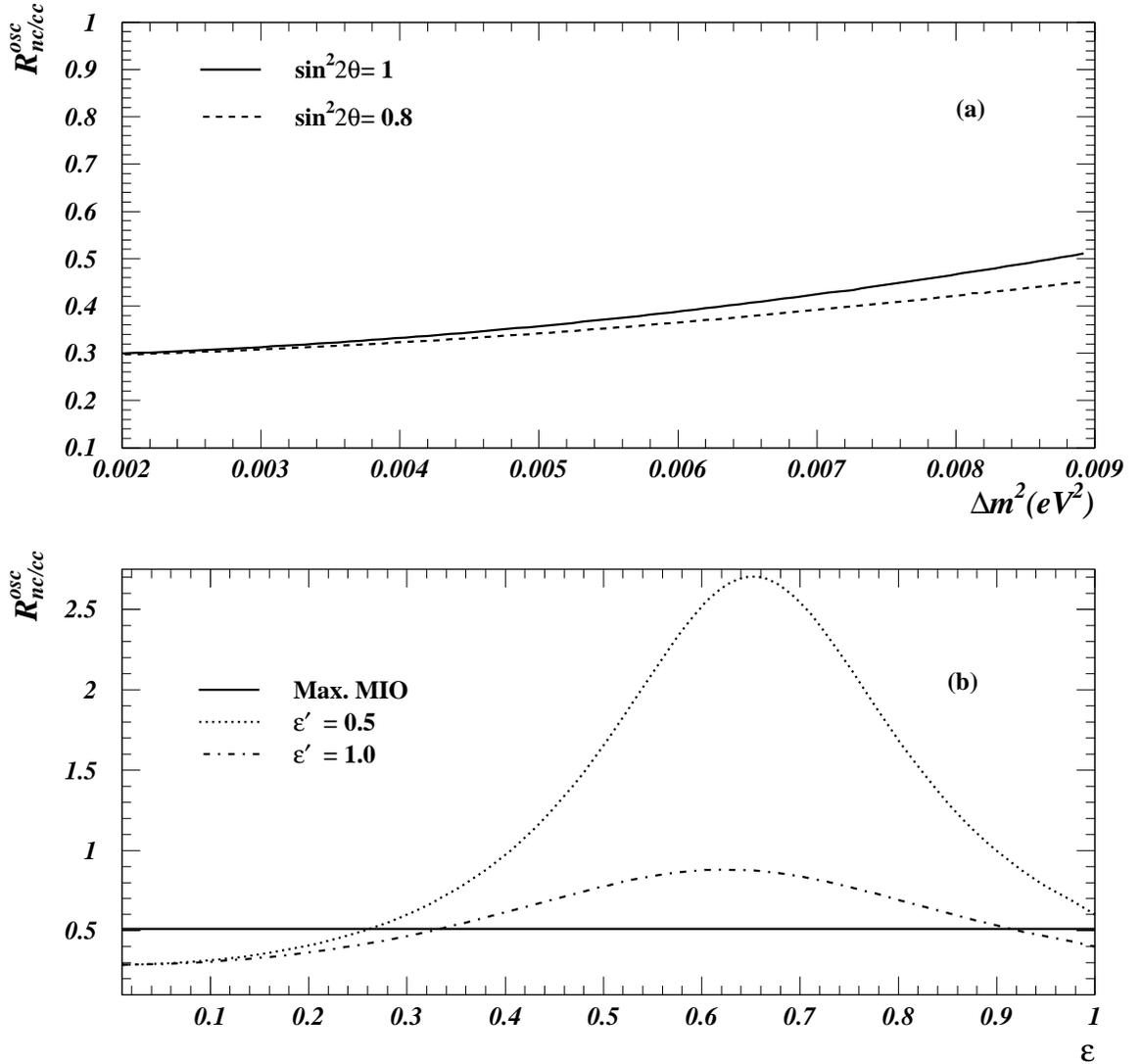,width=\linewidth}}}
\end{center}
\caption{Ratio of nc/cc events at the MINOS far detector for (a) 
MIO as a function of $\Delta m^2$ and (b) FCIO as a function of $\epsilon$.}
\label{fig4}
\end{figure}

For MINOS we have calculated the ratio,
$R_{{\mbox{\scriptsize nc}/{\mbox{\scriptsize cc}}}}^{\mbox{\scriptsize osc}}$,
 which is  given in Eq.~(\ref{ratio1}). This 
is shown in Fig.~\ref{fig4}(a) for the case of MIO. 
Again we have chosen to display this ratio as a function of $\Delta m^2$ 
for the minimum and maximum amplitudes allowed in the 90\% C. L. region 
for Super-Kamiokande solution to the atmospheric neutrino 
problem~\cite{SKconf}. We see that this  mechanism gives rise to 
$R_{{\mbox{\scriptsize nc}/{\mbox{\scriptsize cc}}}}^{\mbox{\scriptsize osc}}$ in the range 0.3-0.52.
The ratio $R_{{\mbox{\scriptsize nc}/{\mbox{\scriptsize cc}}}}^{\mbox{\scriptsize osc}}$  in the hypothesis of FCIO is 
shown in Fig.~\ref{fig4}(b)  as a function of $\epsilon$ for two different 
values of $\epsilon^\prime$. We also show in this figure, as a
reference,  
a flat solid line corresponding to the maximum value of $R_{{\mbox{\scriptsize nc}/{\mbox{\scriptsize cc}}}}^{\mbox{\scriptsize osc}}$  for MIO (0.52). Note that the lowest point 
of the curves in Fig.~\ref{fig4}(a)  is exactly the MINOS sensitivity
limit for the quantity  $R_{{\mbox{\scriptsize nc}/{\mbox{\scriptsize
cc}}}}^{\mbox{\scriptsize osc}}$. From Eq.~(\ref{nn}) one can infer
that for $L=L_{\mbox{\scriptsize K2K}}$, $n_e\sim$ 2.35 mol/cm$^3$
and $C \sim $ 3 we have the maximum of the conversion probability at 
$\epsilon_{*} \sim$ 0.6. The peaks we observe  in Fig.~\ref{fig4}(b)
are due to the fact that we pass by this maximum of the conversion  
probability. Note that even at this point, the contribution due to 
muons generated by tau decays prevents this ratio from going to 
infinity. 

We see clearly from the comparison of Figs.~\ref{fig4}(a) and (b) that 
for $0.35 \lsim \epsilon \lsim 0.9$ it is possible to use 
$R_{{\mbox{\scriptsize nc}/{\mbox{\scriptsize cc}}}}^{\mbox{\scriptsize osc}}$ to distinguish MIO from FCIO, for virtually any 
value of $\epsilon^\prime \leq 1.0$. Nevertheless if the measured nc/cc ratio, 
with its corresponding error, is consistent with the MIO range, 
0.3-0.52, MINOS will be able to rule out a large region in the 
 $\epsilon \times \epsilon^\prime$ plane, but this measurement on its own 
will not be sufficient to completely distinguish  between the two oscillation 
mechanisms since  for $\epsilon \lsim 0.3$ FCIO also predicts a nc/cc ratio 
within this interval. In this case one can try to use the spectral information for 
cc-events and nc-events that will be available at MINOS.

\begin{figure}
\begin{center}
\parbox[c]{6.5in}
{\mbox{\qquad\epsfig{file=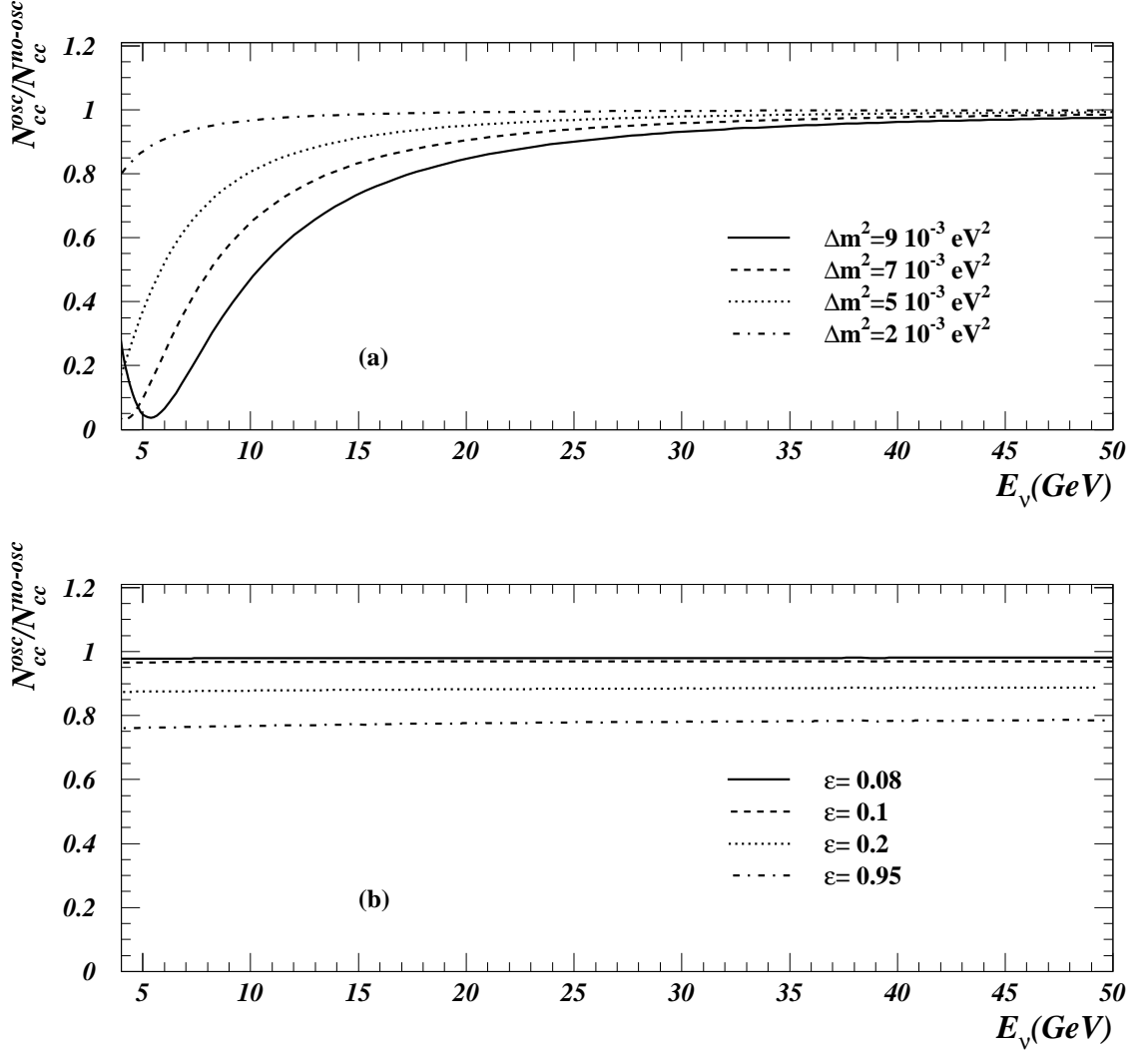,width=\linewidth}}}
\end{center}
\caption{Ratio of the energy spectrum of $\nu_\mu$ cc-events for (a) 
MIO  with $\sin^2 (2\theta)=1.0$ and (b) FCIO with $\epsilon^\prime=1.0$ 
over the expected energy spectrum in the absence of neutrino oscillation for MINOS.}
\label{fig5}
\end{figure}

First one can look for distortions in the energy spectrum for cc-events. 
It can be seen in Fig.~\ref{fig5}(a) that this is very sizable for MIO 
at lower energies as long as  
$\Delta m^2 \gsim 0.004$ eV$^2$, but becomes increasingly difficult to measure 
as  $\Delta m^2$ decreases.  On the other hand for FCIO,  although the 
mechanism is energy independent, we perceive a extremely mild spectral 
distortion in Fig.~\ref{fig5}(b). This happens due to the presence of the second term 
in Eq.~(\ref{ncc}) which carries the cross section energy dependence 
through the $\eta(E_\nu)$ parameter described in Eq.~(\ref{eta}). 
We have plotted in Fig.~\ref{fig5}(b) curves corresponding to four 
different values of $\epsilon$ fixing $\epsilon^\prime=1.0$. 
The ratio $N_{\mbox{\scriptsize cc}}^{\mbox{\scriptsize osc}}/
N_{\mbox{\scriptsize cc}}^{\mbox{\scriptsize no-osc}}$,
for $\epsilon \lsim 0.3$, is completely independence of $\epsilon^\prime$ and is not shown here. 
The four $\epsilon$ values selected  belong to the region in Fig.~\ref{fig4} where the 
$R_{{\mbox{\scriptsize nc}/{\mbox{\scriptsize cc}}}}^{\mbox{\scriptsize osc}}$  
test cannot discriminate between MIO and FCIO.
For $\epsilon \lsim 0.2$ these flat curves may be experimentally indistinguishable  
from the MIO case with $\Delta m^2 \approx 0.002$ eV$^2$. To try to separate  
these   two solutions a precise measurement of the spectrum around 5 GeV 
seems to be crucial.

\begin{figure}[htb]
\begin{center}
\parbox[c]{6.5in}
{\mbox{\qquad\epsfig{file=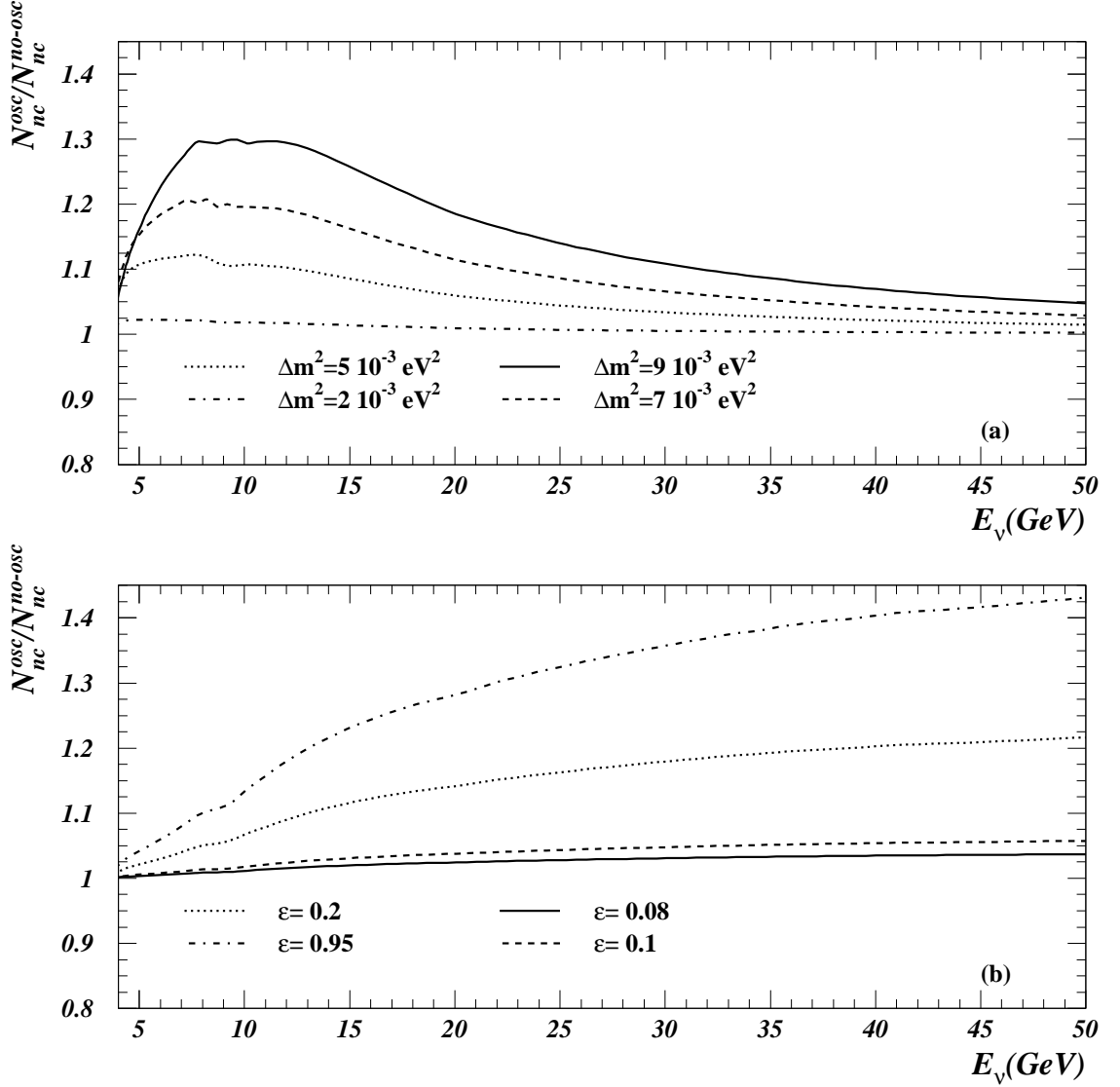,width=\linewidth}}}
\end{center}
\caption{Ratio of the energy spectrum of $\nu_\mu$ nc-events for (a) 
MIO with $\sin^2 (2\theta)=1.0$  and (b) FCIO  with $\epsilon^\prime=1.0$  
over the expected energy spectrum in the absence of neutrino oscillation for MINOS.}
\label{fig6}
\end{figure}

We also can look for distortions in the energy spectrum for nc-events.
As can be seen in Figs.~\ref{fig6}(a)-(b) this presents a quite different behavior 
for MIO and FCIO, but again the case  
$\Delta m^2 \approx 0.002$ eV$^2$ can be hardly distinguished from 
the case $\epsilon \lsim 0.1$.

Now if we compare Fig.~\ref{fig5}(a)-(b) with Figs.~\ref{fig6}(a)-(b) 
we notice a possible smoking gun to discriminate the 
two mechanisms: their different energy dependence for nc- and cc-events. 
MIO in general will cause a pronounced energy distortion for cc-events 
and a  milder energy distortion for nc-events, at low energy and 
decreasing with energy. The FCIO 
mechanism, in contrast,  presents practically no distortion for 
cc-events and a smooth distortion for nc-events, which increases with energy.
Again the latter behavior reflects  basically the energy dependence in 
$\eta(E_\nu)$ for FCIO is an energy independent effect.   
This cross check will be useful for $\Delta m^2 \gsim 0.003$ eV$^2$ and 
$\epsilon \gsim 0.2$.

\section{Conclusions} 
\label{sec:final}

We have studied the possibility of distinguishing mass from flavor changing
induced $\nu_\mu \to \nu_\tau$ conversion in the long baseline experiments 
K2K and MINOS. We have performed a series of estimations of a number of 
observables that will be measured by those two experiments, for these two 
oscillation hypothesis. 

We have calculated the region in the $\epsilon \times \epsilon^\prime$ 
plane that will be observable at K2K and MINOS long baseline experiments.
Although K2K will not be able to completely test the region in 
the $\epsilon \times \epsilon^\prime$ plane which is allowed by the atmospheric 
neutrino FCIO solution given in Ref. \cite{all}, it will be able to cover well the region 
for high values of $\epsilon$. K2K is not very sensitive to $\epsilon^\prime$ 
but is quite sensitive  to $\epsilon \gsim 0.8$. 
MINOS however will be  sensitive to $\epsilon^\prime$ as well as 
to $\epsilon$ and will cover a wider range of these parameters.

>From  Figs.\ref{fig2} we see that if K2K experiment measures a 
averaged survival probability well above 40\% then it can completely 
exclude the FCIO as a possible  solution to the atmospheric neutrino problem. 
But if the measured averaged survival probability  is around or smaller than 
40\%  K2K will have to do very good job in measuring the energy 
 spectrum  of $\nu_\mu$ cc-events to be able to discriminate between FCIO and MIO.

In the case of MINOS, the measurement of the ratio 
$R_{{\mbox{\scriptsize nc}/{\mbox{\scriptsize
cc}}}}^{\mbox{\scriptsize osc}}$ as well as of the spectrum ratios 
 $N_{\mbox{\scriptsize cc}}^{\mbox{\scriptsize
osc}}/N_{\mbox{\scriptsize cc}}^{\mbox{\scriptsize no-osc}}$ and
$N_{\mbox{\scriptsize nc}}^{\mbox{\scriptsize
osc}}/N_{\mbox{\scriptsize nc}}^{\mbox{\scriptsize no-osc}}$ are very powerful tests. 
They will most certainly permit MINOS to explore all the 
region up to $\epsilon \gsim 0.2$.

K2K is now running and they will certainly have results before
MINOS begins to operate. There are three possibilities that one can visualize.
If K2K observes a signal compatible with no-oscillation this implies 
that we are in the worst possible situation for this experiment since  
the average probability observed by K2K would be around or lower than
its  expected experimental sensitivity. 
In principle, this would mean that we are in a bad shape, but a careful analysis
of our results indicates that this is not so. The reason is that this negative 
result would put very strong limits on MIO and FCIO mechanisms that would 
have great consequences on our expectation for MINOS. 
This situation would mean either that $\Delta m^2 <2 \times 10^{-3}$ eV$^2$ for MIO or 
that $\epsilon<0.77$ for FCIO, so while for the MIO mechanism this would imply 
that MINOS should also see a signal compatible with no-oscillation for the FCIO 
mechanism MINOS could still see something.
 Looking at  Fig.~\ref{fig4}(b), we observe that all the 
range $0.32<\epsilon<0.77$ can  be tested  by MINOS independently of the 
value taken by $\epsilon^\prime$. For $\epsilon<0.32$ one 
would still expected some small distortion effect on the MINOS nc-event spectrum. By
comparing the maximal values attained by 
$N_{\mbox{\scriptsize nc}}^{\mbox{\scriptsize osc}}/N_{\mbox{\scriptsize nc}}^{\mbox{\scriptsize no-osc}}$ 
we see that it may be still possible to cover a
extended range of $\epsilon$, depending on $\epsilon^\prime$. 
The more spectacular effect is expected to occur  near the point
$\epsilon \sim  0.6$, where an almost complete conversion of muon neutrinos to tau 
neutrinos occurs.

A second possibility is that K2K measures a positive signal of oscillation. 
It is clear that if K2K measures a  positive signal of 
$\nu_\mu$ flux suppression, which means  $\Delta m^2 > 2 \times 10^{-3}$ eV$^2$ for MIO 
and  $\epsilon > 0.77$ for FCIO,  MINOS will be able to confirm the FCIO solution or discard it 
completely.  We see in Fig.~\ref{fig4}(b) that in this case one should 
expect a striking signal for the ratio 
$R_{{\mbox{\scriptsize nc}/{\mbox{\scriptsize cc}}}}^{\mbox{\scriptsize osc}}$ at MINOS 
if $0.77\leq \epsilon \leq 0.9$, independent of $\epsilon^\prime$.  
Also if $\epsilon>0.9$ the energy spectrum ratios  $N_{\mbox{\scriptsize cc}}^{\mbox{\scriptsize osc}}/N_{\mbox{\scriptsize cc}}^{\mbox{\scriptsize no-osc}}$ and $N_{\mbox{\scriptsize nc}}^{\mbox{\scriptsize osc}}/N_{\mbox{\scriptsize nc}}^{\mbox{\scriptsize no-osc}}$ 
will present for MINOS a very distinct effect for FCIO.

Finally the third possibility. K2K observes a stronger signal 
than the maximum expected  for any of the two
mechanisms, i.e. if the averaged conversion probability is found to be 
much higher than 0.69. In this case neither MIO nor FCIO can explain
such a copious  $\nu_\mu$ disappearance. 
This could be a signal of an additional disappearance channel 
for $\nu_\mu$ and could imply in a not so negligible  contribution of 
$\nu_\mu \to \nu_e$ transitions.

Its clear that the calculations we have performed should be repeated
by the experimentalists, taking into account  efficiencies and other
detector dependent corrections, when this will be known, to compute
more realistic bounds. They should not be very far from
the values given here. We conclude that K2K and MINOS  will be able,
in most envisaged situations, to pin down which 
of the two mechanisms FCIO or MIO really takes place in nature.

\begin{center}
\begin{bf}
Acknowledgments
\end{bf}
\end{center}
We thank Hiroshi Nunokawa and GEFAN for valuable discussions and useful 
comments. We also thank Jorge G. Morfin for  useful correspondence.
This work was supported by Conselho Nacional de Desenvolvimento 
Cient\'{\i}fico e Tecnol\'ogico (CNPq) and by Funda\c{c}\~ao de 
Amparo \`a Pesquisa do Estado de S\~ao Paulo (FAPESP).

\end{document}